\documentclass[conference]{IEEEtran}
\IEEEoverridecommandlockouts
\usepackage{cite}
\usepackage{amsmath,amssymb,amsfonts}
\usepackage{algorithmic}
\usepackage{graphicx}
\usepackage[table]{xcolor}
\usepackage{textcomp}
\usepackage{xcolor}
\def\BibTeX{{\rm B\kern-.05em{\sc i\kern-.025em b}\kern-.08em
    T\kern-.1667em\lower.7ex\hbox{E}\kern-.125emX}}
\begin{document}

\title{A Multi Constrained Transformer-BiLSTM Guided  Network for Automated Sleep Stage Classification from Single-Channel EEG\\
{\footnotesize \textsuperscript{}}
\thanks{}
}

\author{
\IEEEauthorblockN{1\textsuperscript{st} Farhan Sadik}
\IEEEauthorblockA{\textit{Department of EEE} \\
\textit{BUET}\\
Dhaka-1205, Bangladesh \\
1606125@eee.buet.ac.bd}
\and
\IEEEauthorblockN{2\textsuperscript{nd} Md. Tanvir Raihan}
\IEEEauthorblockA{\textit{Department of EEE} \\
\textit{BUET}\\
Dhaka-1205, Bangladesh \\
1606118@eee.buet.ac.bd}
\and
\IEEEauthorblockN{3\textsuperscript{rd} Rifat Bin Rashid}
\IEEEauthorblockA{\textit{Department of EEE} \\
\textit{BUET}\\
Dhaka-1205, Bangladesh \\
1606117@eee.buet.ac.bd}
\and
\IEEEauthorblockN{4\textsuperscript{th} Minhajur Rahman}
\IEEEauthorblockA{\textit{Department of EEE} \\
\textit{BUET}\\
Dhaka-1205, Bangladesh \\
1606126@eee.buet.ac.bd}
\and
\IEEEauthorblockN{5\textsuperscript{th} Sabit Md. Abdal}
\IEEEauthorblockA{\textit{Department of EEE} \\
\textit{BUET}\\
Dhaka-1205, Bangladesh \\
1606119@eee.buet.ac.bd}
\and
\IEEEauthorblockN{6\textsuperscript{th} Shahed Ahmed}
\IEEEauthorblockA{\textit{Department of EEE} \\
\textit{BUET}\\
Dhaka-1205, Bangladesh \\
shahed@eee.buet.ac.bd}
\and
\IEEEauthorblockN{7\textsuperscript{th} Talha Ibn Mahmud}
\IEEEauthorblockA{\textit{Department of EEE} \\
\textit{BUET}\\
Dhaka-1205, Bangladesh \\
talhaibnmahmud30@gmail.com}
}

\maketitle

\begin{abstract}
Sleep stage classification from electroencephalogram (EEG) is significant for the rapid evaluation of sleeping patterns and quality. A novel deep learning architecture, ``DenseRTSleep-II'', is proposed for automatic sleep scoring from single-channel EEG signals. The architecture utilizes the advantages of Convolutional Neural Network (CNN), transformer network, and Bidirectional Long Short Term Memory (BiLSTM) for effective sleep scoring. Moreover, with the addition of a weighted multi-loss scheme, this model is trained more implicitly for vigorous decision-making tasks. Thus, the model generates the most efficient result in the SleepEDFx dataset and outperforms different state-of-the-art (IIT-Net, DeepSleepNet) techniques by a large margin in terms of accuracy, precision, and F1-score.
\end{abstract}

\begin{IEEEkeywords}
Convolutional Neural Network, Deep Learning, Long Short-Term Memory, Sleep stage, Transformer 
\end{IEEEkeywords}

\section{Introduction}
\IEEEPARstart{S}{leep} scoring is an essential tool for sleep disorder diagnosis and treatment \cite{Wulff2010}.
\begin{figure*}[t]
    \centering
    \includegraphics[width=0.9998\linewidth]{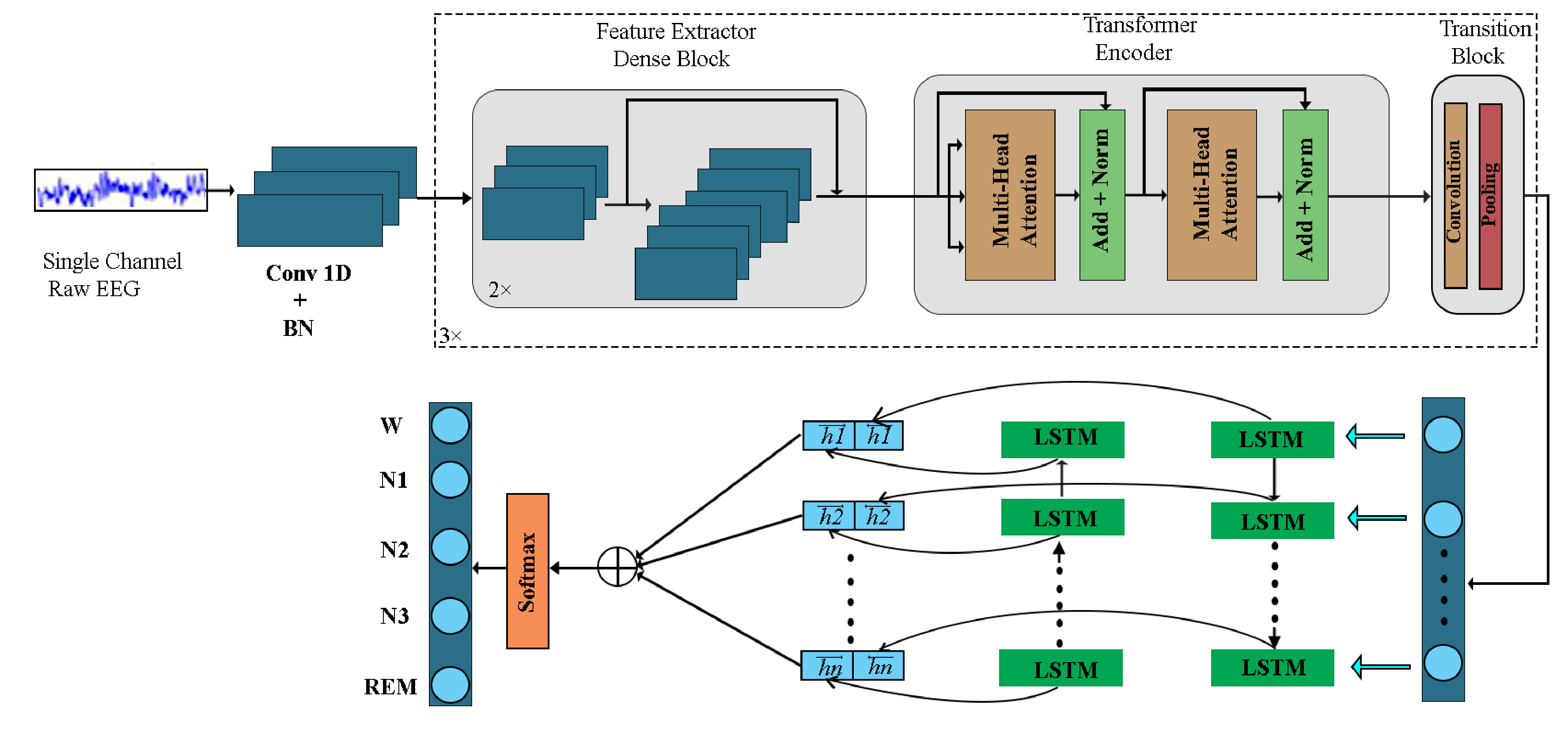}
    \caption{Proposed DenseRTSleep architecture for sleep stage classification.}
    \label{fig:arch}
\end{figure*}
Typical sleep disorders like sleep apnea, narcolepsy, and sleepwalking can be diagnosed via polysomnography (PSG) which is the gold standard in this regard. PSG records the biosignals of body functions such as brain activity (electroencephalogram, EEG), eye movement (electrooculogram, EOG), heart rhythm (electrocardiogram, ECG), and muscle activity of the chin, face, or limbs (electromyogram, EMG). Following the acquisition of recording, these signals are analyzed by trained professionals, who split the PSG data into 20-s or 30-s segments (an ``epoch") along with the corresponding sleep stage labeling. The sleep stages can be termed as Wakefulness (W), Rapid Eye Movement (REM), and Non-REM (NREM) following the Rechtschaffen and Kales (R\&K) rules \cite{rechtschaffen1968manual}.
According to the AASM rules, NREM is again divided into three stages, namely S1, S2, and S3, or N1, N2, and N3. The professionals have to examine all the epochs and label their corresponding sleep stages to draw a whole-night hypnogram demonstrating the sleep stages as a function of sleep time. This type of manual sleep scoring is certainly very tiring and time-consuming \cite{Stepnowsky2013}. Hence, automatic sleep stage classification systems are very much a necessity to assist sleep specialists.
Many conventional machine learning methods to classify sleep stages from EEG signals are widely used. These methods mostly comprise two steps, namely feature extraction and sleep stage classification. For the first step, the design and extraction of various features from time or frequency or both domains \cite{Hassan2016} are performed. Additionally, several feature selection algorithms are often utilized for a further selection of the most discriminating features. On the other hand, for the second step, the selected features are then fed into the conventional machine
learning models, namely, Support Vector Machines (SVM) \cite{Faust2019}, Random Forest (RF) \cite{Li2018}, or ensemble learning-based classifiers \cite{Hassan2016} for sleep stage classification. However, these methods require specific domain knowledge to extract the best representative features without any feedback mechanism.
Recently, deep learning has been extensively used in different areas and performed quite superiorly to other conventional machine learning models without much domain knowledge which motivates researchers to exploit deep learning techniques for automatic sleep stages classification. Several studies have utilized Convolutional Neural Networks (CNNs) \cite{Li2021} for this task. Though the CNN models of these studies have achieved good performance for sleep stage classification, most of them are not able to effectively model the temporal dependencies among the EEG data as CNN is best known for extracting effective spatial features only. So, Recurrent Neural Network-Long Short-Term Memory (RNN-LSTM) was proposed to capture temporal dependencies in time-series EEG data. For example, \cite{Michielli2019} used a cascaded RNN-LSTM architecture to classify sleep stages fr. Some researchers combined CNN with RNN-LSTM to explore both the special and temporal features \cite{Seo2020, Supratak2017}. However, RNN-LSTMs also have limitations due to their recurrent nature i.e., they usually have high model complexity and it is thus difficult to train them. So, RNN-LSTM along with any deep network requires a very large training time. 
Again, \cite{9697331} proposed a transformer, a sequential model solely based on self-attention, and was able to achieve similar state-of-the-art results. Although CNN, RNN-LSTM, and transformer have individual strong points they have some shortcomings such as high memory requirements, and poor performance in the small class. Hence finding a shallow architecture that efficiently incorporates all of these networks by minimizing their limitations is still a challenging task. 
To tackle the above difficulties, we propose a transformer-BiLSTM guided convolutional neural network called ``DenseRTSleep-II'' for automatic sleep stage classification from single-channel EEG signal. This shallow architecture consists of three building blocks for spatial feature apprehension namely the feature extractor block, transformer encoder block, and the transition block.  The feature extractor block consists of two consecutive dense blocks and extracts efficient features from the EEG signals. The transformer encoder block having a multi-head attention mechanism encodes these features in different representations and passes this representation to the transition block.
At the final stage of the architecture, two BiLSTM layers are included to preserve the temporal contexts of the sequence. Moreover, a novel multi-loss scheme is proposed for efficient decision-making at the prediction stage. Through comprehensive experimentation, the effectiveness of different parts of the ``DenseRTSleep-II'' architecture is ensured. The novel contribution of this transformer-BiLSTM guided shallow network is that it uses the transformer network as an encoder for enhancing the efficiency of the temporal context analyzer, LSTM along with a multi-loss scheme which effectively increases the learning ability. 

\section{Methodology}
In this paper, we propose a transformer-BiLSTM guided shallow convolutional neural network for sleep stage classification from single-channel EEG. The network consists of a transformer encoder enclosed by a feature extractor dense block and a transition block. Two BiLSTMs in both directions are incorporated to apprehend the temporal contexts. The performance of the DenseRTSleep-I (Dense Recurrent network along with a Transformer encoder for Sleep stage classification) is enhanced by utilizing a multi-loss scheme that draws effective boundaries between discriminative features (DenseRTSleep-II) whereas the previous methods utilize only categorical cross-entropy as loss function. In order to analyze the performance of each element of the network, we name the sub-networks as DenseSleep, DenseRNNSleep, DenseRTSleep-I, and DenseRTSleep-II which is summarized in table \ref{Ablation}. The color ``green" denotes the utilized features for a particular model. 
 
\begin {table} [t] \label{Ablation} 
\centering
\caption{Categorization of individual network elements \\ Green colored cell indicates the utilized block for a particular model}
\footnotesize
\resizebox{.45\textwidth}{!}{
\begin{tabular}{ |p{1.8cm}|p{1.8cm}|p{1.8 cm}|p{1.8 cm}|  } 
\hline
\textbf{DenseSleep} & \textbf{DenseRNNSleep} & \textbf{DenseRTSleep-I} & \textbf{DenseRTSleep-II} \\
\hline
\cellcolor[HTML]{93F588} Feature extractor & \cellcolor[HTML]{93F588} Feature extractor &\cellcolor[HTML]{93F588}  Feature extractor & \cellcolor[HTML]{93F588} Feature extractor \\
\hline
\cellcolor[HTML]{93F588} Transition block & \cellcolor[HTML]{93F588} Transition block  & \cellcolor[HTML]{93F588} Transition block & \cellcolor[HTML]{93F588} Transition block \\
\hline
BiLSTM & \cellcolor[HTML]{93F588} BiLSTM & \cellcolor[HTML]{93F588} BiLSTM & \cellcolor[HTML]{93F588} BiLSTM \\
\hline
Transformer encoder  & Transformer encoder & \cellcolor[HTML]{93F588} Transformer encoder & \cellcolor[HTML]{93F588} Transformer encoder \\
\hline
Multi-loss scheme & Multi-loss scheme & Multi-loss scheme &\cellcolor[HTML]{93F588}  Multi-loss scheme\\
\hline
\end{tabular}
}
\end{table}

The proposed DenseRTSleep model is shown in Fig. \ref{fig:arch}.

\subsection{Feature Extractor Block}
The DenseSleep model involves a feature extractor block and a transition block repeated three times before entering a softmax layer. The feature extractor block consists of a dense block and connects it to the softmax layer with a transition block inspired by DenseNets \cite{huang2017densely}. Although the original DenseNets perform supervised classification on 2D image data, the proposed DenseSleep model utilizes a custom 1D implementation for sequential data. The feature extractor block consists of batch normalization (BN) followed by a convolution through an activation function RELU and the operations being repeated two times. Skip connections are added to encounter the vanishing gradient problem to keep consistency with the densenets. The transition block consists of convolution and pooling operations to sequentially transfer the low-level features into high-level features.

\subsection{Temporal Context Encoder}
DenseRNNSleep includes two BiLSTMs \cite{greff2016lstm} with the DenseSleep with hidden states of dimension u = 128. The BiLSTM performs certain operations on the feature space with the previous hidden state thus achieving feature representation in both forward $\overrightarrow{\bold{h_{t}}}$ and backward direction $\overleftarrow {\bold{h_{t}}}$ at time step t. It can be formulated as

\begin{equation}
\overrightarrow{\bold{h_{t}}} = \overrightarrow{\mathrm{LSTM}}(\bold{f_{t}},\overrightarrow{\bold{h_{t-1}}})
\end{equation}
\begin{equation}
\overleftarrow{\bold{h_{t}}} = \overleftarrow {\mathrm{LSTM}}(\bold{f_{t}},\overleftarrow{ \bold{h_{t+1}}})
\end{equation}

\subsection{Transformer Encoder}
A transformer encoder is integrated with the DenseRNNSleep in between the feature extractor and the transition block to translate the representation of the features (DenseRTSleep-I). A multi-head self-attention mechanism (MHA) is utilized in the transformer encoder block \cite{vaswani2017attention}. The attention mechanism enables the model to split the sequence into different feature subspaces and attain long-term dependencies with any sequence length. The MHA takes three duplicate input features $\bold{X} = {x_{1},x_{2},...,x_{N}}$, where N is the number of features vectors. The attention mechanism can be formulated as 
\begin{equation}
\mathrm{ATT}(\hat{\bold{X}},\hat{\bold{X}},\hat{\bold{X}})=\mathrm{Softmax}(\frac{\hat{\bold{X}}.\hat{\bold{X}^{T}}}{\sqrt{d}}).\hat{\bold{X}}
\end{equation}
where $\hat{\bold{X}}$ is the output of a convolution operation performed on $\bold{X}$. MHA subdivides the $\hat{\bold{X}}$ into H subspaces i.e., $\hat{\bold{X}} = {X^{1},...,X^{H}}$.  H denotes the number of heads. For the proposed method, H = 3. In each subspace h, the MHA is formed as
\begin{equation}
\bold{A^{h}} = \mathrm{ATT}(\hat{\bold{X^{h}}},\hat{\bold{X^{h}}},\hat{\bold{X^{h}}}) 
\end{equation}

\begin{equation}
\mathrm{MHA}(\hat{\bold{X^{h}}},\hat{\bold{X^{h}}},\hat{\bold{X^{h}}})= \mathrm{Concat}(\bold{A^{1}},...,\bold{A^{H}})
\end{equation}

The DenseRTSleep-I utilizes the post-BN layer i.e., the normalization layer is used after the MHA layer. The transformer encoder block is repeated three times with the feature extractor block and the transition block as depicted in Fig. \ref{fig:arch}.

\subsection{Multi loss Scheme}

In this work, we propose a weighted multi-loss scheme to supervise the learning task of the neural network model. Here a linear combination of the contrastive loss \cite{wang2021understanding} and Kullback–Leibler divergence (KL-divergence) \cite{yu2013kl} loss is utilized along with the categorical cross-entropy loss. The intuition behind the multi-loss scheme is that it clusters the familiar features together and repulses the other features. In short, it increases the distance between the features of the different classes and makes it an easy learning task to draw the decision boundary. The total loss for the problem is the weighted linear combination of the loss functions described. Combining the multi-loss strategy with the previous DenseRTSleep-I, we propose DenseRTSleep-II for the effective classification of sleep stages from single-channel EEG data. The loss functions are formulated below:

\begin{equation}
L_{\text {cross-entropy }}\left(y_{\text {true }}, y_{\text {pred }}\right)=\sum y_{\text {true }}* \\ 
\log \left(y_{\text {predicted }}\right)\\ 
\end{equation}

\begin{equation}
\begin{array}{r}
L_{K L-\text { divergence }}\left(y_{\text {true }}, y_{\text {pred }}\right)=\sum y_{\text {true }} * \\
\log \left(y_{\text {true }} / y_{\text {predicted }}\right)
\end{array}
\end{equation}

where \emph{D} is the Euclidian distance between the true class and the predicted class and ``$\mu$" reinforce a constraint on the loss i.e., if two features are dissimilar then the deviation should be at least some margin $\mu$, otherwise, a loss will occur.

\begin{equation}
\begin{array}{r}
L_{\text {contrastive-loss }}\left(y_{\text {true }}, y_{\text {pred }}\right)=\sum y_{\text {true }} * D^2+ \\
\left(1-y_{\text {true }}\right) * \max (\mu-D, 0)
\end{array}
\end{equation}

\begin{equation}
\begin{array}{r}L_{\text {Total }}=L_{\text {cross-entropy }} +\alpha L_{\text {contrastive-loss }} \\ +\beta L_{K L-\text { divergence }}\end{array}
\end{equation}

Trial and error simulated experimentations indicate the best performance achieved for the ``DenseRTSleep-II" is with $\alpha$ = 0.1 and $\beta$ = 0.9.

\section{Experimental Setup}

\begin{figure}[htbp]
\centering
\includegraphics[width=0.60\linewidth]{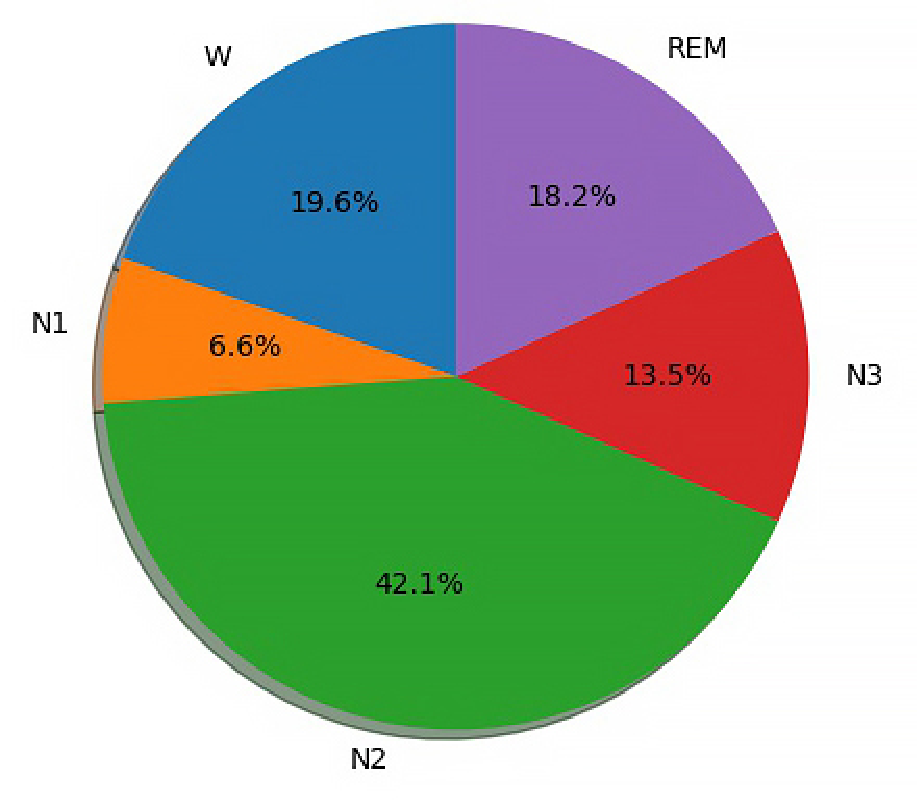}
\caption{Data distribution for 20 patients.}
\label{fig:dd}
\end{figure}

\subsection{Dataset}
To evaluate the performance of the proposed models, SleepEDFx, a publicly available dataset containing PSG records and their corresponding sleep stages labeled by experts is used\cite{goldberger2000physiobank}. It contains two types of PSG records i.e., SC for 20 healthy subjects without sleep-related disorders and ST for 22 subjects of a study on Temazepam effects on sleep. Each record consists of two-channel EEGs from the Fpz-Cz and Pz-Oz channels, a single-channel EOG, and a single-channel EMG. Each half-minute epoch is labeled as one of eight classes (W, REM, N1, N2, N3, N4, MOVEMENT, UNKNOWN) according to R\&K rules. In this work, The Fpz-Cz channel is used. We split the dataset into three partitions where 16 patients were used for training and validation (12 and 4) and 4 patients were used for testing purposes. It is apparent that the dataset is highly imbalanced and the number of samples for the ``N2" class is very large compared to the other classes as shown in Fig. \ref{fig:dd}. 

\subsection{Evaluation Metrics}
Evaluation metrics used to evaluate the performance of the results are single-class precision, single-class recall, single-class F1-score, and overall accuracy. All of them are defined using True Positive (TP), True Negative (TN), False Positive (FP), and False Negative (FN) detections from the confusion matrix of a particular model. Single-class evaluation is considered a binary classification with fixating the particular class as positive and keeping all others as negative. 

Precision and recall primarily focus on the number of TP detections of a particular class which can be formulated as

\begin{align}
   \bold{Precision} = \frac{\sum \bold{TP}}{\sum (\bold{TP}+\bold{FP})}\\
   \bold{Recall} = \frac{\sum \bold{TP}}{\sum (\bold{TP}+\bold{FN})}
\end{align}

The F1-score represents the harmonic mean of precision and recall as follows:
\begin{align}
   \bold{F1-score} = 2*\frac{\bold{Precision}*\bold{Recall}}{\bold{Precision} + \bold{Recall}}
\end{align}

Finally, accuracy is simply the ratio of all true detections to all the other observations which can be formulated as

\begin{align}
   \bold{Accuracy} = \frac{\sum(\bold{TP}+\bold{TN})}{\sum(\bold{TP+TN+FP+FN})}
\end{align}
\\

\section{Result and Discussion}
\begin{table}[t]
\centering
\caption{Performance Analysis of DenseRTSleep-II model}
\footnotesize
\label{Result4}

\begin{tabular}{|c|c|c|c|c|c|c|c|c|} 
\hline
\textbf{ }&\multicolumn{5}{|c|}{\textbf{Predicted}} &\multicolumn{3}{|c|}{\textbf{Per Class Metrices}}\\
\hline

\hline
\textbf{ Actual}& \textbf{W} & \textbf{N1}
& \textbf{N2}& \textbf{N3}& \textbf{REM} & \textbf{Prec} & \textbf{Rec} & \textbf{F1-score} \\ 
\hline
{W}  
                       & 1800&	110&	61&	5&	150	&0.8699&	0.8466&	0.8581
          \\ 

                       \hline

{N1}                            
                       & 140&	160&	71	&1&	170	&0.2862&	0.2952&	0.2906
          \\ 
                       \hline
{N2}                            
                 &59	&140	&3300	&110&	270&	0.8371&	0.8507&	0.8438
          \\ 
                       \hline
{N3}                            
            & 17&	9&	270	&1100&	0	&0.9038	&0.7879	&0.8419
          \\ 
                       \hline 

{REM}                            
                    &  53&	140&	240&	1&	1300&	0.6878&	0.7497&	0.7174
          \\ 
                       \hline
\end{tabular}
\end{table}

\begin{table}[t]
\centering
\caption{Comparison performance of different methods}
\footnotesize
\label{Result5}

\begin{tabular}{|c|c|c|c|c|} 

\hline
\text{Model}& \textbf{Precision} & \textbf{Recall} & \textbf{F1-score}
& \textbf{Accuracy}\\ 
\hline
IIT-NET \cite{Seo2020}    &70.17\%&	65.98\%&	67.59\%	&75.77\% 
          \\ 

                       \hline
IT-NET \cite{Seo2020}                   
                       & 69.48\%	&65.21\%&	66.63\%&	75.16\% 
          \\ 
                       \hline
DeepSleepNet \cite{Supratak2017}                      
                 &68.93\% &	\textbf{73.07}\%	&69.76\%&	74.86\% 
          \\ 
                       \hline
\textbf{DenseRTSleep-II}                       
            & \textbf{71.70\%}& 70.60\%&	\textbf{71.04\%}&	\textbf{79.16\%} 
          \\ 
                       \hline

\end{tabular}
\end{table}

A noteworthy fact in the training dataset is the imbalance distribution of the N2 case. And the true detection of the N1 cases is poor among all of the state-of-the-art methods. In the ablation study, four variations were tested to ensure the performance of the proposed method. 
At first, the performance of the feature extraction block followed by a transition block (DenseSleep) was evaluated. The overall accuracy obtained here was 79.45\% which is a significant improvement over the state-of-the-art methods.

\begin{figure}[t]
    \centering
    \includegraphics[width=0.9998\linewidth]{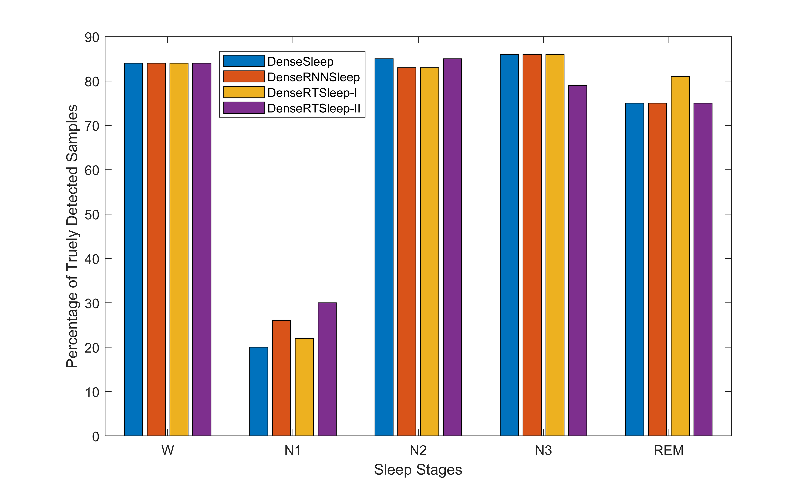}
    \caption{Performance analysis of individual components of DenseRTSleep-II.}
    \label{fig:analysis}
\end{figure}

We can see a clear improvement in the N1 classes for DenseRNNSleep, where 140 samples were correctly detected as shown in Fig. \ref{fig:analysis}. The previous model (DenseSleep) was able to detect 110 correct samples. It is apparent that the LSTM block successfully captures the temporal features of the EEG sequences. Then integrating the transformer encoder block between the feature extractor block and the transition block (DenseRTSleep-I), significant improvement in the overall accuracy was seen as depicted in Fig. \ref{fig:analysis}. 
And finally, the result of our proposed model, DenseRTSleep–II, where multi-loss constraints are integrated with the previous model and a huge improvement in the case of N1 detection is seen, where 160 samples are correctly detected and the result is also somewhat balanced between the true classes than the previous results as shown in table \ref{Result4}.

Finally, the overall comparison matrix of our proposed method with state of art methods is presented in Table \ref{Result5}. The overall highest accuracy among the state-of-the-art methods is about 4\% less than our proposed method. It is to be noted that IIT-net (intra- and inter-epoch) and IT-net (intra-epoch) were able to detect 130 and 110 N1 classes whereas our proposed method was able to detect 160 true N1 cases. Here a clear improvement in the overall accuracy, precision, and F1 score of our proposed method over the state-of-the-art methods is observed. The number of parameters of this model is 7,49,913 which is significantly lower than the state-of-the-art deeper models available in the literature thus making the DenseRTSleep-II a shallow efficient network for fast sleep stage classification. 

In the result section of our paper, we presented a comprehensive analysis of the performance of our proposed DenseRTSleep-II model for sleep stage classification. However, we would like to clarify the differences in reported accuracy between our work and previous methods, specifically IIT-NET and DeepSleepNet. The variation in reported accuracy values can be attributed to differences in experimental settings and data splits. We used a distinct test set for our evaluation, consisting of 16 training and 4 test patients, while IIT-NET and other previous methods typically averaged their results over test sets with 19 training and 1 test patient. This divergence in test set composition can lead to variations in reported accuracy due to the inherent variability in patient data. To ensure a fair and rigorous comparison, we did not rely solely on the reported values from the papers. Instead, we implemented these methods in our own environment, using the same data splits and experimental conditions for all methods, including our proposed DenseRTSleep-II. This approach allowed us to directly compare the performance of our method with IIT-NET and other state-of-the-art methods under consistent circumstances.

\section{Conclusion}
In this paper, a novel transformer-BiLSTM guided shallow neural network, DenseRTSleep-II is proposed to exploit both the spatial and the temporal dependencies of a sequence along with a multi-loss scheme as a prediction enhancer for end-to-end sleep stage classification from single-channel EEG. The transformer block is used in between the feature extractor and the transition block, constructing a different representation of the sequence that flows to the forward network. Finally, the BiLSTMs create representative temporal contexts from the sequence of the features and a softmax layer predicts the sleep stages. Experimental results show the effectiveness of each element and the improvement over the state-of-the-art methods available in the literature which makes DenseRTSleep-II, an eligible candidate for fast and accurate prediction of sleep stages from single-channel EEG. 

\bibliographystyle{IEEEtran}
\bibliography{eeg.bib}

\begin{thebibliography}{10}
\providecommand{\url}[1]{#1}
\csname url@samestyle\endcsname
\providecommand{\newblock}{\relax}
\providecommand{\bibinfo}[2]{#2}
\providecommand{\BIBentrySTDinterwordspacing}{\spaceskip=0pt\relax}
\providecommand{\BIBentryALTinterwordstretchfactor}{4}
\providecommand{\BIBentryALTinterwordspacing}{\spaceskip=\fontdimen2\font plus
\BIBentryALTinterwordstretchfactor\fontdimen3\font minus \fontdimen4\font\relax}
\providecommand{\BIBforeignlanguage}[2]{{%
\expandafter\ifx\csname l@#1\endcsname\relax
\typeout{** WARNING: IEEEtran.bst: No hyphenation pattern has been}%
\typeout{** loaded for the language `#1'. Using the pattern for}%
\typeout{** the default language instead.}%
\else
\language=\csname l@#1\endcsname
\fi
#2}}
\providecommand{\BIBdecl}{\relax}
\BIBdecl

\bibitem{Wulff2010}
\BIBentryALTinterwordspacing
K.~Wulff, S.~Gatti, J.~G. Wettstein, and R.~G. Foster, ``Sleep and circadian rhythm disruption in psychiatric and neurodegenerative disease,'' \emph{Nature Reviews Neuroscience}, vol.~11, no.~8, pp. 589--599, Jul. 2010. [Online]. Available: \url{https://doi.org/10.1038/nrn2868}
\BIBentrySTDinterwordspacing

\bibitem{rechtschaffen1968manual}
A.~Rechtschaffen, ``A manual for standardized terminology, techniques and scoring system for sleep stages in human subjects,'' \emph{Brain information service}, 1968.

\bibitem{Stepnowsky2013}
\BIBentryALTinterwordspacing
C.~Stepnowsky, D.~Levendowski, D.~Popovic, I.~Ayappa, and D.~M. Rapoport, ``Scoring accuracy of automated sleep staging from a bipolar electroocular recording compared to manual scoring by multiple raters,'' \emph{Sleep Medicine}, vol.~14, no.~11, pp. 1199--1207, Nov. 2013. [Online]. Available: \url{https://doi.org/10.1016/j.sleep.2013.04.022}
\BIBentrySTDinterwordspacing

\bibitem{Hassan2016}
\BIBentryALTinterwordspacing
A.~R. Hassan and M.~I.~H. Bhuiyan, ``Computer-aided sleep staging using complete ensemble empirical mode decomposition with adaptive noise and bootstrap aggregating,'' \emph{Biomedical Signal Processing and Control}, vol.~24, pp. 1--10, Feb. 2016. [Online]. Available: \url{https://doi.org/10.1016/j.bspc.2015.09.002}
\BIBentrySTDinterwordspacing

\bibitem{Faust2019}
\BIBentryALTinterwordspacing
O.~Faust, H.~Razaghi, R.~Barika, E.~J. Ciaccio, and U.~R. Acharya, ``A review of automated sleep stage scoring based on physiological signals for the new millennia,'' \emph{Computer Methods and Programs in Biomedicine}, vol. 176, pp. 81--91, Jul. 2019. [Online]. Available: \url{https://doi.org/10.1016/j.cmpb.2019.04.032}
\BIBentrySTDinterwordspacing

\bibitem{Li2018}
\BIBentryALTinterwordspacing
X.~Li, L.~Cui, S.~Tao, J.~Chen, X.~Zhang, and G.-Q. Zhang, ``{HyCLASSS}: A hybrid classifier for automatic sleep stage scoring,'' \emph{{IEEE} Journal of Biomedical and Health Informatics}, vol.~22, no.~2, pp. 375--385, Mar. 2018. [Online]. Available: \url{https://doi.org/10.1109/jbhi.2017.2668993}
\BIBentrySTDinterwordspacing

\bibitem{Li2021}
\BIBentryALTinterwordspacing
F.~Li, R.~Yan, R.~Mahini, L.~Wei, Z.~Wang, K.~Mathiak, R.~Liu, and F.~Cong, ``End-to-end sleep staging using convolutional neural network in raw single-channel {EEG},'' \emph{Biomedical Signal Processing and Control}, vol.~63, p. 102203, Jan. 2021. [Online]. Available: \url{https://doi.org/10.1016/j.bspc.2020.102203}
\BIBentrySTDinterwordspacing

\bibitem{Michielli2019}
\BIBentryALTinterwordspacing
N.~Michielli, U.~R. Acharya, and F.~Molinari, ``Cascaded {LSTM} recurrent neural network for automated sleep stage classification using single-channel {EEG} signals,'' \emph{Computers in Biology and Medicine}, vol. 106, pp. 71--81, Mar. 2019. [Online]. Available: \url{https://doi.org/10.1016/j.compbiomed.2019.01.013}
\BIBentrySTDinterwordspacing

\bibitem{Seo2020}
\BIBentryALTinterwordspacing
H.~Seo, S.~Back, S.~Lee, D.~Park, T.~Kim, and K.~Lee, ``Intra- and inter-epoch temporal context network ({IITNet}) using sub-epoch features for automatic sleep scoring on raw single-channel {EEG},'' \emph{Biomedical Signal Processing and Control}, vol.~61, p. 102037, Aug. 2020. [Online]. Available: \url{https://doi.org/10.1016/j.bspc.2020.102037}
\BIBentrySTDinterwordspacing

\bibitem{Supratak2017}
\BIBentryALTinterwordspacing
A.~Supratak, H.~Dong, C.~Wu, and Y.~Guo, ``{DeepSleepNet}: A model for automatic sleep stage scoring based on raw single-channel {EEG},'' \emph{{IEEE} Transactions on Neural Systems and Rehabilitation Engineering}, vol.~25, no.~11, pp. 1998--2008, Nov. 2017. [Online]. Available: \url{https://doi.org/10.1109/tnsre.2017.2721116}
\BIBentrySTDinterwordspacing

\bibitem{9697331}
H.~Phan, K.~B. Mikkelsen, O.~Chen, P.~Koch, A.~Mertins, and M.~De~Vos, ``Sleeptransformer: Automatic sleep staging with interpretability and uncertainty quantification,'' \emph{IEEE Transactions on Biomedical Engineering}, pp. 1--1, 2022.

\bibitem{huang2017densely}
G.~Huang, Z.~Liu, L.~Van Der~Maaten, and K.~Q. Weinberger, ``Densely connected convolutional networks,'' in \emph{Proceedings of the IEEE conference on computer vision and pattern recognition}, 2017, pp. 4700--4708.

\bibitem{greff2016lstm}
K.~Greff, R.~K. Srivastava, J.~Koutn{\'\i}k, B.~R. Steunebrink, and J.~Schmidhuber, ``Lstm: A search space odyssey,'' \emph{IEEE transactions on neural networks and learning systems}, vol.~28, no.~10, pp. 2222--2232, 2016.

\bibitem{vaswani2017attention}
A.~Vaswani, N.~Shazeer, N.~Parmar, J.~Uszkoreit, L.~Jones, A.~N. Gomez, {\L}.~Kaiser, and I.~Polosukhin, ``Attention is all you need,'' \emph{Advances in neural information processing systems}, vol.~30, 2017.

\bibitem{wang2021understanding}
F.~Wang and H.~Liu, ``Understanding the behaviour of contrastive loss,'' in \emph{Proceedings of the IEEE/CVF conference on computer vision and pattern recognition}, 2021, pp. 2495--2504.

\bibitem{yu2013kl}
D.~Yu, K.~Yao, H.~Su, G.~Li, and F.~Seide, ``Kl-divergence regularized deep neural network adaptation for improved large vocabulary speech recognition,'' in \emph{2013 IEEE International Conference on Acoustics, Speech and Signal Processing}.\hskip 1em plus 0.5em minus 0.4em\relax IEEE, 2013, pp. 7893--7897.

\bibitem{goldberger2000physiobank}
A.~L. Goldberger, L.~A. Amaral, L.~Glass, J.~M. Hausdorff, P.~C. Ivanov, R.~G. Mark, J.~E. Mietus, G.~B. Moody, C.-K. Peng, and H.~E. Stanley, ``Physiobank, physiotoolkit, and physionet: components of a new research resource for complex physiologic signals,'' \emph{circulation}, vol. 101, no.~23, pp. e215--e220, 2000.

\end{thebibliography}

\end{document}